\documentclass[nonacm,sigplan,natbib=true]{acmart} % For LaTeX2e
\AtBeginDocument{%
  \providecommand\BibTeX{{%
    \normalfont B\kern-0.5em{\scshape i\kern-0.25em b}\kern-0.8em\TeX}}}
\usepackage{tikz}
\usepackage{forest}
\usepackage{array}
\usepackage{subfiles}
\usepackage{multirow}
\usepackage{tabularray}
\usepackage{colortbl}
\usepackage{makecell}
\usepackage{hhline}
\begin{document}

\begin{abstract}
We present a novel evaluation framework for representation bias in latent factor recommendation algorithms. Our framework introduces the concept of attribute association bias in recommendations allowing practitioners to explore how recommendation systems can introduce or amplify stakeholder representation harm. Attribute association bias (AAB) occurs when sensitive attributes become semantically captured or entangled in the trained recommendation latent space. This bias can result in the recommender reinforcing harmful stereotypes, which may result in downstream representation harms to system consumer and provider stakeholders. Latent factor recommendation (LFR) models are at risk of experiencing attribute association bias due to their ability to entangle explicit and implicit attributes into the trained latent space. 
Understanding this phenomenon is essential due to the increasingly common use of entity vectors as attributes in downstream components in hybrid industry recommendation systems. We provide practitioners with a framework for executing disaggregated evaluations of AAB within broader algorithmic auditing frameworks. Inspired by research in natural language processing (NLP) observing gender bias in word embeddings, our framework introduces AAB evaluation methods specifically for recommendation entity vectors. We present four evaluation strategies for sensitive AAB in LFR models: attribute bias vector creation, attribute association bias metrics, classification for explaining bias, and latent space visualization. We demonstrate the utility of our framework by evaluating user gender AAB regarding podcast genres with an industry case study of a production-level DNN recommendation model. We uncover significant levels of user gender AAB when user gender is used and removed as a model feature during training to mitigate explicit user gender bias, pointing to the potential for systematic bias in LFR model outputs. Additionally, we discuss best practices, learnings, and future directions for evaluating AAB in practice.
\end{abstract}
%% Full title
\title{Evaluation Framework for Understanding Sensitive Attribute Association Bias in Latent Factor Recommendation Algorithms}
\author{Lex Beattie}
\email{lex@spotify.com}
\affiliation{%
  \institution{Spotify}
      \country{USA}
}
\author{Isabel Corpus}
\email{isabelc@spotify.com}
\affiliation{%
  \institution{Spotify}
      \country{USA}
}

\author{Lucy H. Lin}
\email{lucyl@spotify.com}
\affiliation{%
  \institution{Spotify}
      \country{USA}
}

\author{Praveen Ravichandran}
\email{praveenr@spotify.com}
\affiliation{%
  \institution{Spotify}
      \country{USA}
}
\renewcommand{\shortauthors}{Beattie et al.}
\maketitle
\section{Introduction}
\label{sec:intro}
Latent factor recommendation (LFR) algorithms have become fundamental to industry recommendation settings \cite{burke2007hybrid, amatriain2015recommender}. These recommendation algorithms, such as collaborative filtering and deep learning, provide predictions of engagement and embedded vector representations of users and items. The resulting trained vector representations can capture entity relationships and characteristics in a condensed dimensional space and allow for comparisons between different entity vectors in the trained latent semantic space. 

It has been demonstrated that user and item attributes can become entangled when leveraging these algorithms, resulting in feature duplication and bias amplification \cite{zhang2020content}. This algorithmic outcome can result in lower and less robust recommendation quality \cite{zhang2020content}. Research seeking to reduce this type of attribute disentanglement has become increasingly prevalent and showcases favorable results when targeting exposure bias attributed to item attributes, popularity, or user behavior \cite{zhang2020content, zheng2021disentangling, nema2021disentangling, divitiisfashion2022, zhao2022popularity, wang2022causal}. However, the research primarily focuses on intrinsic mitigation techniques and increasing recommendation performance but does not always provide evaluation techniques to understand how the bias may be captured explicitly or implicitly within the latent space. Attribute disentanglement traditionally requires attributes to be independent and explicitly used in order to implement disentanglement methods. This common requirement results in disentanglement evaluation methods failing to address situations where attributes show interdependence with one another or present themselves implicitly in results. This stipulation hinders processes for identifying systematic bias, such as gender or racial bias, that can be interdependent with other attributes or be implicitly captured by behavior in the recommendation scenario. 

Other research concerning this type of bias in representation learning has shown that systematic bias can occur due to the nature of latent factor algorithms, as previously found in research exploring systematic gender bias in natural language processing (NLP) \cite{gonen2019lipstick}. Leveraging the framework introduced in this paper, we demonstrate that this risk can also occur within the outputs of latent factor recommendation. Given the popularity of LFR models in industry systems and the use of their outputs as downstream features in hybrid or multi-component recommendation systems, it is paramount that practitioners understand and can evaluate attribute association bias to reduce the risk of introducing or reinforcing representation bias in their recommendation systems \cite{amatriain2015recommender}.

In other areas of representation learning, such as NLP, implicit or systematic bias has been shown to result in downstream representation harms (e.g., translation systems being more likely to generate masculine pronouns when referring to stereotypically-male occupations \cite{savoldi2021mt}). NLP researchers have studied this type of bias by evaluating and mitigating  \textit{gender association bias} \cite{bolukbasi2016man, basta2019evaluating, caliskan2017}. Even though association bias evaluation has been a focus in other areas of representation learning, it remains largely unstudied for recommendation systems \cite{ekstrand_fairness_ias}. In this work, we close this gap by presenting a framework of methodologies for evaluating \textit{attribute association bias} resulting from LFR algorithms. Unlike previous work focusing on gender, our framework is designed to be attribute agnostic, thus the name attribute association bias.
Attribute association bias is present when entity embeddings showcase significant levels of association with specific types of explicit or implicit entity attributes. For example, while users can be explicitly labeled by gender, pieces of content cannot be gendered. However, due to the potential for attribute entanglement, pieces of content can show measurable levels of implicit association with a gender attribute.

Understanding how attribute association bias can become entangled within the trained latent space is vital in industry recommendation settings because of the popularity of latent factor algorithms and the common practice of leveraging embedding outputs in hybrid recommendation systems \cite{burke2007hybrid, amatriain2015recommender}. 
In these industry recommendation system settings, learned vectors may be used in unrelated model systems to create product predictions based on user and item representations. If sensitive attribute bias is encoded into the vector, it plausibly can be repeated and amplified when said vectors are used as features in other models.
Ignoring this type of bias puts practitioners at risk of unknowingly amplifying stereotypes and representative harm within their recommendation systems. For example, \citet{amatriain2015recommender} described how matrix factorization algorithms could be combined with ``traditional neighborhood-based approaches'' to create a recommendation system for Netflix. This combination consisted of LFR embeddings ranked based on neighborhood-based algorithms to produce final recommendations for consumption \cite{amatriain2015recommender}. If certain sets of user or item vectors were closely associated with a sensitive attribute, the resulting attribute association bias could affect final rankings in the KNN outputs due to groups of vectors forming stereotyped clusters due to this association bias.

Our proposed evaluation methods for understanding attribute association bias in LFR models can be coupled with bias mitigation techniques to first understand if the case requires bias mitigation and subsequently determine if the mitigation technique was successful. 
This enables greater transparency when it would otherwise be challenging to interpret the causal effects of attributes on recommendation representations. 
Evaluation frameworks, such as the one presented in our paper, are essential to practitioners, allowing them to thoroughly investigate the level of bias in their system before experimenting with and completing expensive mitigation techniques \cite{richardson2021_rubric}.

Our work provides a practical guide for evaluating attribute association bias in trained vector embeddings. We introduce recommendation entity-specific attribute association vector directions, bias metrics, and evaluation techniques inspired by gender association bias NLP research. Our methods account for differences between recommendation system and NLP representation embeddings and are designed to provide flexibility for evaluation of binary attributes beyond gender bias. Our framework is model agnostic concerning the type of LFR algorithm. Additionally, our methods allow for the analysis of attribute association bias for any type of binary attribute.
In addition to introducing the methodology, we implement these methods on an industry case study to demonstrate how these techniques can be implemented in practice and best practices for interpreting evaluation results. 

To the best of our knowledge, we present one of the first evaluation frameworks for addressing attribute association bias (as a type of representation bias). Our case study also presents one of the first quantitative analyses of user gender bias in latent factor recommendations for podcasts. Our work lowers the barrier to analyzing attribute representation bias in LFR algorithms and opens the door to disentanglement of dependent attributes and implicit representation bias in recommendation systems.

Our contributions include:
\begin{itemize}
    \item Definition of attribute association bias within the context of latent factor recommendation algorithms (\S\ref{sec:problem}).
    \item Evaluation methodologies and metrics for analyzing attribute association bias between recommendation entity embeddings (\S\ref{sec:methods}).
    \item Recommendations for implementation in practice by showcasing techniques with a case study observing user gender attribute association bias resulting from a podcast recommendation model (\S\ref{sec:casestudy}-\ref{sec:results}).
    \item Discussion of limitations of this approach and future directions (\S\ref{sec:discussion}-\ref{sec:limits}).
\end{itemize}

Throughout the paper, we leverage language such as stereotypes, bias, and harm. When referring to bias, we are discussing algorithmic or qualitative statistically skewed results found in experimental or evaluation settings which can produce harm \cite{crawford_keynote}. We refer to stereotypes as a ``product of biases" often held at the societal level, which may or may not be supported in experimental settings \cite{alexander1992makes}. Stereotyping algorithmic harm occurs when it is produced by algorithmic bias \cite{crawford_keynote}. This type of harm can be seen as ``representative" harm due to it reinforcing ``the subordination of some groups along the lines of identity" \cite{crawford_keynote}. Our framework, and presented case study, looks to quantitatively evaluate algorithmic bias (attribute association bias) which signals potential for reinforcing stereotypes thus resulting in downstream representative harm.

%The paper will be structured as follows, section 2 will frame our contributions into the broader research area of fairness and bias in recommendations and natural language processing. Section 3 will define our problem statement and how our contributions will benefit the information retrieval research community. Section 4 will present our methodologies for evaluating attribute association bias. Section 5 focuses on implementing said methodologies on an industry-specific case study. Section 6 will discuss the results from our evaluation and the learnings from implementing our proposed techniques. Finally, before the paper's conclusion, the last two sections will focus on discussing our findings concerning previous work in representation harms and systematic bias in word embeddings, limitations of our research, and future directions to improve our framework for evaluating attribute association bias.
\section{Related Work}
\label{sec:related}
Our evaluation framework for analyzing attribute association bias contributes to the research area of fairness and bias in recommendations. Additionally, we build upon seminal work in addressing gender attribute association bias in natural language processing. In this section, we provide an overview of relevant key findings and research.

\subsection{Fairness and Bias in Recommendations} 
Understanding how to define and evaluate fairness and bias of recommendation systems has quickly grown into a seminal area of information retrieval research. Various types of bias related to recommendation systems have been defined and studied in academic and industry settings. Researchers often target studying bias relating to harms of allocation, unequal distribution of exposure, or attention of recommendations within the system \cite{ekstrand_fairness_ias}. Allocative, or distributional, harms have been studied by evaluating and mitigating biases such as popularity, exposure, ranking (or pairwise), and gender bias \cite{abdollahpouri_reducing_2019, abdollahpouri_user-centered_2021, abdollahpouri_value-aware_2018, geyik_fairness-aware_2019, dash2021umpire, mehrotra_towards_2018, ekstrand_gender_dist}. A recent literature review by \citet{ekstrand_fairness_ias} notes that representational harms can also be studied in recommendation systems but focuses on representation in terms of the provider and how stakeholders view their distribution within the system, not their numerical representation as vector outputs of a recommendation system. 

In this paper, we explore representation harm in terms of association bias, more commonly studied in NLP, to understand how stereotypes can become encoded into the latent embedding space \cite{ekstrand_fairness_ias, bolukbasi2016man}. Previously, in NLP research, this bias was referred to as gender association bias \cite{ bolukbasi2016man}. Our proposed framework and methodologies build upon previous NLP methods for gender association bias to analyze association bias agnostic of the type of bias in recommendation settings. Our modifications account for distinct differences between NLP and LFR models and generalize beyond gender when evaluating association bias.

Our proposed methodologies are designed to help uncover implicit or explicit attribute association bias that can affect latent factor recommendation models and may unknowingly amplify stereotypes or contribute to downstream allocation harms. Our work evaluating attribute association bias presents a new direction in exploring bias in recommendations by addressing the current gap in quantifying representation bias in the vector representations leveraged and outputted by a recommendation system.

In addition to studying a form of bias not often addressed in recommendation system research, we also present metrics that address concepts not currently captured by recommendation fairness research. Current metrics target distributional harms resulting from the biases mentioned above by measuring the equity of recommendation results through \textit{accuracy} based, \textit{pairwise} based, or \textit{causal} based methods \cite{verma2020facets, ekstrand2022fairness}. Instead of focusing on distributional outcomes, our techniques focus on increasing the transparency of entity relationships embedded into the latent space by evaluating entity bias associations. These relationships can be studied regarding individual and group fairness, where group fairness focuses on providing similar outcomes across groups of entities \cite{dwork_fairness_2011}, whereas individual fairness specifies that similar entities should receive similar outcomes \cite{dwork_fairness_2011}. Our framework focuses on evaluating the group fairness of attribute association bias to understand if one group of entities experiences more stereotyped encodings than another.

\subsubsection{Gender Bias}
Our case study (\S\ref{sec:casestudy}) presents an offline evaluation of attribute association bias relating to user gender stereotypes, specifically how recommendations of specific pieces or types of content may vary according to a user's gender. Beyond the scope of this work, offline evaluation of user gender bias in recommendations has been largely unexplored due to the difficulty in obtaining the user's gender for analysis. User gender bias evaluation is often only available to industry practitioners, given their ability to collect and analyze user attribute data. However, results are only sometimes shared due to the sensitive nature of this work. We hope that by sharing our findings concerning user gender bias, we increase transparency within this space and encourage other industry researchers and practitioners to follow suit, thus allowing a greater ability for industry and academia to collaborate, address, and mitigate these kinds of challenges.

Given this, past research on evaluating gender bias in recommendations has more frequently focused on allocation harms occurring from provider or creator gender bias in the recommendation system. For example, \citet{ekstrand_gender_dist} explored the effects of author gender on book recommendations by exploring item recommendation distributions according to author gender, showing that recommendation algorithms could potentially recommend specific authors in a gender-biased fashion. \citet{shakespeare2020exploring} evaluated the extent to which artist gender bias is present within music recommendations. \citet{ferrarobreakloop2021} conducted experiments studying how collaborative filtering performs when specifically accounting for an artist's gender in music recommendation. Our work differentiates from past research by exploring representational harms of \emph{user} gender resulting from LFR models. 

Even though quantifying user gender bias in media recommendation systems remains rare, researchers have produced work leveraging qualitative user studies to evaluate gender differences in media preference. For example,
\citet{millar2008selective} conducted a user study of young female and male adults to evaluate "gender differences in artist preferences". \citet{berkers2012gendered} conducted an offline analysis of gender preferences in the Last.FM dataset, showcasing how data can capture gendered listening patterns in music.

\subsection{Quantifying Gender Associations in Natural Language Processing Representations}

Our proposed evaluation framework for measuring entity bias associations in latent recommendation model embeddings is inspired by natural language processing (NLP) methods that attempt to measure binary gender bias in word embeddings --- e.g., associations between gender-neutral words (like ``scientist'' or ``nurse'') and words indicative of a specific gender (like ``man'' or ``woman''). Past work has identified gender biases in pretrained static word embeddings \cite[\emph{inter alia}]{bolukbasi2016man, caliskan2017}, contextual word embeddings from large language models \cite{basta2019evaluating, tan2019assessing, zhao2019gender, bhardwaj2021investigating}, and embeddings of larger linguistic units like sentences \cite{may2019measuring, liang2020towards}. Because pretrained word \& sentence embeddings are widely used as input for many NLP models, there is potential for biases in embeddings to be propagated or amplified in downstream text classification and generation tasks \cite{zhao2019gender,orgad2022gender,savoldi2021mt}. % (TODO@lucyl: footnote the field's tendency to assess English models/social contexts and binary gender; see if there's a good citation for bias in embeddings used for retrieval; describe why these cannot just be applied directly to recsys somewhere)

In the recommendation system setting, we consider the analogous possibility of sensitive feature associations in latent entity embeddings due to the possibility of propagation in downstream models \cite{burke2007hybrid}. However, unlike analysis of gender in relation to words, recommendation entities do not necessarily have natural contrastive pairings by attribute. We address this gap in \S\ref{sec:methods}.

\section{Problem Statement}
\label{sec:problem}
Disentangled latent factor recommendation research has become increasingly popular as LFR algorithms have been shown to entangle model attributes in their resulting trained user and item embeddings, leading to unstable and inaccurate recommendation outputs \cite{zhang2020content, zheng2021disentangling, nema2021disentangling, wang2022causal}.
However, most of this research is outcome-focused, providing mitigation methods for improving performance but not addressing the potential for representation bias in the latent space.
As a result, few existing evaluation techniques analyze how attributes are explicitly (due to distinct use as a model attribute) or implicitly captured in the recommendation latent space. 
For those that do exist, the metrics focus on evaluating disentanglement levels for explicitly used and independent model attributes, instead of investigating possible implicit bias associations between entity vectors and sensitive attributes or systematic bias captured within the latent space \cite{nema2021disentangling}.
Even though latent representation bias has become a well-studied phenomenon in other types of representation learning, such as natural language and image processing, it remains relatively under-examined compared to the large amounts of research concerning exposure and popularity bias \cite{ekstrand_fairness_ias}.

The work presented in this paper looks to close the current research gap concerning evaluating representation bias in LFR algorithms by providing a framework for evaluating \textit{attribute association bias}. 
Identifying potential attribute association bias encoded into user and item (entity) embeddings is essential when they become downstream features in hybrid multi-stage recommendation systems, often encountered in industry settings \cite{burke2007hybrid, amatriain2015recommender}. 
Evaluating the \textit{compositional fairness} of these systems, or the potential for bias from one component to amplify into downstream components, is challenging if one does not understand how this type of bias initially occurs within the system component \cite{wang2021practical}.
Understanding the current state of bias is imperative when auditing and investigating the system prior to mitigation in practice \cite{beattie2022challenges}. 
Our proposed methods seek to lower the barrier for practitioners and researchers wishing to understand how attribute association bias can infiltrate their recommendation systems. These evaluation techniques will enable practitioners to more accurately scope what attributes to disentangle in the mitigation and provide baselines for deeming the mitigation successful. 

We apply these methods to an industry case study to assess user gender attribute association bias in a LFR model for podcast recommendations. Prior research primarily has focused on evaluating provider gender bias due to the lack of publicly available data on user gender bias; to the best of our knowledge, our work provides one of the first looks into quantifying user gender bias in podcast recommendations. We hope that our observations help other industry practitioners to evaluate user gender and other sensitive attribute association bias in their systems, provide quantitative insights into podcast listening beyond earlier qualitative user studies, and encourage future discussion and greater transparency of sensitive topics within industry systems.

\section{Methodology}
\label{sec:methods}
Our methodology is intentionally designed to support implementation in practice, specifically within broader algorithmic auditing frameworks. Broad system-level frameworks, such as  ``SMACTR" or ``SIIM", provide essential structure to approaching the ambiguous problem space of auditing for harms and bias in industrial systems \cite{raji2020closing, beattie2022challenges}. %\citet{raji2020closing} introduced the seminal auditing framework of "SMACTR", standing for "Scoping, Mapping, Artifact Collection, Testing, and Reflection." SMACTR is viewed as a framework for a large-scale audit, explicitly accounting for "procedures and
%documentation, as well as considering system outputs" \cite{barocas2021designing}. 
Our proposed framework falls more in line with the idea of "disaggregated evaluations," which is targeted by the SIIM framework or the "Testing" step of SMACTR, where the focus lies on analyzing AI outputs for harm or bias \cite{raji2020closing, beattie2022challenges, barocas2021designing}.

We present our framework for a disaggregated evaluation of latent factor recommendation algorithms leveraging the SIIM framework for analyzing recommendation bias in practice \cite{beattie2022challenges}. This framework consists of four steps: scope, identify, implement, and monitor and flag. The first step, scope, addresses the problem of determining ``what'' to analyze. For example, what sensitive attribute should be analyzed in our LFR model vector output? The second step, identify, focuses on determining the best-suited methodologies for said analysis based on the scope and outputs of the system. The third step, implement, represents the time dedicated to conducting the analysis and determining how to manipulate the data to leverage identified methods. Finally, the fourth step, monitor and flag, answers the vital question of if significant levels of bias exist within the system. 

Our methodology framework provides practitioners with guidance to analyze attribute association bias systematically with the SIIM framework. We do this by specifically focusing on how to scope the attribute to be measured for bias, identifying and implementing our methodologies to determine the existence of bias, and then finally, flagging for significant results and significant changes in bias after mitigation. The following three sections capture these four steps: scope, identify \& implement, and flag. We combine identify and implement to guide when to leverage a method in conjunction with quantitative instructions for implementation. We focus on flagging over monitoring since our framework addresses ad hoc testing for significance rather than setting baselines for long-term systematic monitoring. It is an essential, yet challenging, task to standardize the creation of baselines for monitoring, since determining levels of harm is highly context specific \cite{madaio2020co, madaio2022assessing}. Providing practical guidance for setting baselines is out of the scope of this paper but can be seen as an impactful direction for future research.

\subsection{Scope}
When approaching bias evaluation, practitioners must first scope what they wish to measure \cite{beattie2022challenges}. In the case of attribute association bias, one must define which attribute needs to be targeted during the evaluation.
For our proposed evaluation framework, an attribute can refer to any entity characteristic defined by a binary relationship differentiating two entity groups (e.g., ``male'' and ``female'' for binary gender). Scoping the attribute will involve defining the two characteristically opposing groups of entity embeddings for calculating the attribute association bias. In addition to defining the attribute and its respective entity groups, one must determine what groups of entity embeddings should be tested for attribute association bias. For example, if one defines user gender as the attribute, the practitioner would need to determine how to group items as test entities for analysis. However, if the practitioner wishes to observe artist gender bias, they would need to define groups of user entities for their evaluation. The test entity embeddings should be considered based on their perceived risk of attribute association bias. For example, if a group of entities is historically stereotyped, that group's embeddings could be a candidate for attribute association bias testing. 
\subsection{Identify \& Implement}
The next step of our framework focuses on identifying and implementing methods for evaluating attribute association bias. First, we introduce and provide implementation instructions for exploring attribute association bias. In addition to introducing these methods, we provide guidance for identifying which methods to use during the evaluation to help practitioners find the correct methods for their specific use case.

We present four categories of evaluation methods for exploring attribute association bias in latent factor recommendation algorithms: (1) latent space visualization, (2) bias directions, (3) bias evaluation metrics, and (4) classification for explaining bias. We introduce these methods in order of the type of analysis the practitioner wishes to conduct, from initial exploration to targeted measurement for determining mitigation needs. Thus, we provide support for evaluating bias across different phases of analysis while addressing bias and harm within one's recommendation system.

When first exploring the existence of attribute association bias, we suggest implementing latent space visualization and bias direction methods. These two methods may alert practitioners to the existence of significant levels of attribute association bias. Latent space visualization provides an easily interpretable view of the attribute. It can signal when more analysis is needed, but it should not be used as a quantitative measurement for the existence of bias. The next group of methods, bias directions, provides quantitative means for determining if a significant relationship exists between the entities and the attribute. One can leverage these methods to answer the question: ``does attribute association bias exist?'' 

If a clear attribute relationship has been identified, one may investigate the problem further by measuring the level of bias present. This type of level setting and direct measurement can be done by implementing bias evaluation metrics and classification techniques to test for significant levels of attribute association bias and create statistical baselines for evaluating if mitigations are successful. Practitioners can implement these methods to address the question: ``how strong is the level of attribute association bias in my system?''

When describing implementation details, we refer to the attribute-defining entity sets as \(A\) and \(B\). 
Each entity, \(a \in A\) and \(b \in B\), is assigned a binary label representing the attribute, with the label of set \(A\) being one and that of \(B\) being zero, or vice versa. 
These two entity sets can be considered opposing if their labeled attribute is mutually exclusive. 
Entity sets used to test for attribute association bias will be referred to as \(E\) and \(P\). 
It is assumed that one entity set is hypothesized to show heavier stereotyping towards one of the opposing attribute entity sets.

%Describing how relationships exist within an n-dimensional space is not always intuitive. In order to help showcase how our methods capture these complex relationships, we will leverage a 2D example to serve as our demonstration case to accompany our details below. The entity groups are generated randomly and split across the x-axis, with entities from A and E existing in a two-dimensional vector space ranging from 0 < x < 100 and 0 < y < 100. We also normalize this space so that each vector reflects a one-unit vector. Finally, we set our attribute entity sets E and P in two opposite parts of the example latent space. This creates a clear linear relationship between the "attribute" entity sets, E and P. We designate two scenarios to show how one may wish to leverage particular methods given their scenario.

%\begin{figure}
 %   \includegraphics[scale=0.4]{main_v2/methodology/implement/Screenshot 2023-03-16 at 1.57.05 PM.png}
%    \caption{Visualization of the 2D example used to demonstrate methodologies in our proposed framework.}
 %   \label{fig:my_label}
%\end{figure}
\subsubsection{Latent Space Visualization}
\label{subsec:biasviz}
Latent space visualization has been shown to be a valuable technique for qualitatively evaluating entity relationships \cite{liu2019latent}. This method can be an effective first step before implementing quantitative methodologies for bias measurement.

In past work, dimension reduction techniques such as principal component analysis (PCA) and t-distributed stochastic neighbor embedding (t-SNE \cite{van2008visualizing}) have been used to visualize latent variables in high dimensional data. In one such example, \citet{gonen2019lipstick} use t-SNE to visualize gender clusters in word embeddings. We adapt this methodology to understand the latent representations of attribute associations in our feature embeddings. 

However, unlike earlier work, we suggest using PCA instead of t-SNE to visualize feature embeddings when evaluating grouping. t-SNE representations privilege representations of local, rather than global, structures in high-dimensional data, thereby obfuscating the relationships between attribute clusters in visualizations \cite{van2008visualizing}. Since we recommend dimensionality reduction visualization techniques as a qualitative method to understand the extent of clustering around attributes, t-SNE's distortion of global geometry is consequential. We also use PCA to apply learned mapping to features not used in training, which would not have been possible with t-SNE. We recommend using PCA to visualize the first and second principal components, which capture the directions of most variability in feature embeddings \cite{bolukbasi2016man}. 
\subsubsection{Bias Directions}
\label{subsec:biasvectors}
Calculating attribute bias directions can serve as another method for exploring the existence of bias in one's system. 
These \emph{attribute association bias direction vectors} represent how the attribute is distinguished as a vector direction between $A$ and $B$ within the trained latent space. These vectors can be used for: (a) exploring individual entities by identifying users or items whose embeddings have high similarity with a particular attribute association bias vector for further examination; (b) comparing recommendation systems by using the bias vectors to calculate association bias metrics (\S\ref{subsec:biasmetrics}) for each system; and (c) exploring classification scenarios (\S\ref{subsec:biasclassify}).

We present three methods for computing attribute association bias direction vectors: centroid difference, SVC vector direction, and PCA vector direction. Unlike related work in NLP association bias research (e.g., \cite{bolukbasi2016man}), the centroid difference and SVC vector direction calculations do not require practitioners to have distinct representation embedding pairings between entities in $A$ and $B$, making them suitable for recommendation systems and data.
% now discussed in intro to section : These two methods instead assumes that an attribute has (or is assigned) two opposing \emph{sets} of entity vectors (notated here as $A$ and $B$) that capture the attribute's values. For example, if one wishes to calculate a bias vector for binary gender, the practitioner could group users or items as male or female.

\paragraph{Centroid Difference}
The simplest method for computing an attribute's association bias vector is to take the difference between the centroid of $A$ and the centroid of $B$ (also referred to as attribute vector mapping \cite{liu2019latent}). This method is best used for capturing differences in \emph{average} attribute behavior. The centroid method is the most readily interpretable of the three due to its simple calculation and thus serves as a good starting point for exploring the attribute space. However, it is essential to note that this method tends to be more conservative in estimating bias due to variance being averaged out in the process. It may not adequately capture significant nuances in behavior within the space, and other direction techniques may be required to reflect more complex attribute bias behavior.

\paragraph{SVC Vector Direction}
Our second approach computes the association bias vector using parameters from a linear support classification (SVC) model trained to predict the attribute. We draw inspiration for this technique from past NLP research that trained SVC models to predict grammatical gender in word embeddings \cite{omrani2022measuring, zhou2019examining}. The entity vectors and labels in sets $A$ and $B$ are used as training data for the SVC model. The attribute bias direction is created from the final attribute layer of the model to capture the subspace representing significant attribute meaning. The selection and assignment of entities to $A$ and $B$ can substantially impact the computed bias direction; in our case study (\S\ref{sec:casestudy}), we compare bias directions computed on random samples of users versus most stereotypically-gendered ones. This direction methodology is best used to capture more distinct nuances of attribute bias that may be lost when entity vectors are averaged in the centroid difference method.

%From observing bias directions for our 2D example, we found that the SVC vector direction served as a good middle ground for measuring attribute association bias. It was able to capture a more significant bias direction than that of the centroid direction without requiring the creation of a seed set to leverage the PCA method.

\paragraph{PCA Vector Direction}
The final method calculates the bias direction by using the general methodology introduced by \citet{bolukbasi2016man}, which is based on conducting principal component analysis (PCA) on parallel attribute pair vectors $\{(a_0, b_0), \ldots, (a_{|A|}, b_{|B|})\}$. The final attribute bias direction is the first eigenvector of these vectors, capturing the majority of the variance found describing the group of vector pairs. Similar to the methods above, two groups of opposing vectors must be defined to create the final pairing of vectors. However, unlike the two methods above, implementing PCA for vector direction creation requires distinct attribute pair vectors. This better enables visualization for transparency, but should only be used if the practitioner is confident in their entity pairings for defining the attribute. Therefore, this may not be a good starting point for bias exploration. We show this caveat in our case study by presenting the downfalls of randomly selecting attribute entity pairs for creating a PCA vector direction.

%When leveraging this method for our 2D example, we looked at two possible cases. First, hand choosing the most "extreme" points to calculate the difference vector. Our other method consisted of randomly drawing points from the attribute set for the difference vector. We found that hand picking the seed set resulted in a more significant bias direction while the randomly chosen bias direction resulted in a less accurate direction vector for relaying attribute association bias. This supports our recommendation that this method should solely be used when the practitioner is confident in their seed set. 
\subsubsection{Bias Evaluation Metrics}
\label{subsec:biasmetrics}
We propose two metric methods for capturing attribute association bias culminating from LFR algorithms. Two NLP techniques inspire these evaluation methods for evaluating latent gender bias in word embeddings: Word Embedding Association Test (WEAT \cite{caliskan2017}) and Relational Inner Product Association (RIPA \cite{ethayarajh2019understanding}). We chose these methods based on their acceptance within the NLP community as reliable metrics for measuring bias in vector representations of words \cite{du2021assessing, zhang2020robustness}. In this section, we present adaptations of WEAT and RIPA for use in recommendation settings.

\paragraph{Entity Attribute Association Metrics \& Test}
This set of metrics, inspired by WEAT \cite{caliskan2017}, can be used to understand how attribute association bias  manifests in user-user and user-item comparisons by computing vector similarity between entities of interest and members of the two attribute groups defined previously (\(A\) and \(B\)). These metrics require two sets of users or items to evaluate the attribute association bias (\(E\) and \(P\)), where one entity set is hypothesized to show heavier stereotyping than the other. There are three interrelated metrics: entity attribute association (EAA), group entity attribute association (GEAA), and differential entity attribute association (DEAA).

EAA measures the attribute association bias for a single entity \(\varepsilon \in \{E \cup P\}\), calculated as the difference in mean cosine similarity of \(\varepsilon\) to attribute entities in \(A\) and \(B\). Positive EAA scores represent a higher association with attribute \(A\) while negative scores signal higher association with attribute \(B\):
\begin{equation} \label{eaa}
EAA (\varepsilon, A, B) = \frac{\sum_{a\in A}cos(\varepsilon , a)}{\left | A \right |} - \frac{\sum_{b\in B}cos(\varepsilon , b)}{\left | B \right |} 
\end{equation}

GEAA is the sum of all entity attribute association scores for a set of entities (\(E\) or \(P\)):

\begin{equation} \label{geaa}
GEAA (E, A, B) = \sum_{\varepsilon\in E} EAA (\varepsilon, A, B)
\end{equation}

Finally, DEAA acts as the test statistic for permutation testing by measuring the scale in the difference between the GEAA of \(E\) and \(P\). Positive DEAA scores signal that entities in \(E\) show more association with attribute \(A\) than entities in \(P\) and vice versa for negative scores:

\begin{equation} \label{deaa}
DEAA (E, P, A, B) = GEAA(E, A, B) - GEAA(P, A, B)
\end{equation}

We use permutation testing to evaluate if there is a significant difference in how the test entity sets relate to the attribute entity sets. Additionally, we adopt the calculation for effect size presented by \citet{caliskan2017} to evaluate the normalized separation between EAA distributions of the test entity sets:

\begin{equation} \label{effect}
\frac{\frac{GEAA(E, A, B))}{\left | E \right |} - \frac{GEAA(P, A, B)}{\left | P \right |}}{
stddev(GEAA(E\bigcup P, A, B))
}
\end{equation}

\paragraph{Recommendation Relational Inner Product Association (R-RIPA)}
We also provide a metric, R-RIPA, that is similar to the prior metrics, but parameterized by a user-defined attribute bias direction. This provides more flexibility for the practitioner to use computed attribute association bias vectors based on SVC and PCA, or other user-defined attribute directions in general. Additionally, this metric may be more robust to fluctuations or outliers that can affect metrics heavily reliant on group averages over entities.

We base this metric on RIPA \cite{ethayarajh2019understanding}, which is calculated with a relation vector representing the first principal component of the difference between word pairings in an attribute-defining set.
We modify RIPA to require a relation vector \(\psi\) that represents a user-defined attribute bias direction between $A$ and $B$. R-RIPA for an entity set $E$ is computed as:

\mathchardef\mhyphen="2D
\begin{equation} \label{rripa}
R\mhyphen RIPA (E, \psi) = \frac{\sum_{\varepsilon\in E}cos(\varepsilon , \psi)}{\left | E \right |}
\end{equation}

The effect size for R-RIPA between entity sets \(E\) and \(P\) is then:

\begin{equation} \label{rripa_effect}
\frac{R\mhyphen RIPA (E, \psi) - R\mhyphen RIPA (P, \psi)}{
stddev\{cos(\varepsilon, \psi) \mid E\bigcup P \}
}
\end{equation}

\subsubsection{Classification for Explaining Bias}
\label{subsec:biasclassify}
Past NLP research has used classification models to show how heavily word embeddings capture bias and demonstrate possible downstream effects, particularly along the lines of binary gender \cite{gonen2019lipstick,basta2019evaluating}. We propose similar use of classifiers on entity representations to explore attribute association bias in recommendation settings. 

More specifically, we propose training a classifier on user or item embeddings and their target attribute, meaning the model is trained on entity sets \(A\) and \(B\) and their associated attribute labels. This classifier can then be leveraged to explore how attribute bias and stereotypes are captured within the trained latent space, e.g., by comparing predictions of new entities not in \(A\) and \(B\). This method is especially advantageous when assessing the potential for amplifying representation harm when using item or user embeddings in models downstream from the original recommendation system; we demonstrate its utility in our case study (\S\ref{sec:casestudy}).

\subsection{Flag}
When implementing an audit or evaluation of bias, one of the most important steps is determining if bias exists, requires mitigation, or has fluctuated significantly due to a system change or recently completed mitigation \cite{beattie2022challenges}. We introduce significance testing methods for the quantitative methods described in the previous section (bias directions, bias evaluation metrics, and classification results). Note that we do not suggest specific baselines for flagging required mitigation since baselines should be determined on a case-by-case basis \cite{beattie2022challenges}. This section does not address interpreting visualizations given their qualitative nature, but in general, the practitioner should look for clear delineation between entity attribute and test sets. A widening or narrowing of the linear separation would signal changes in latent visualization.

\subsubsection{Bias Directions}
Unlike testing significance for metrics, testing the significance of a latent direction requires validating that the direction captures attribute behavior. We aim to determine that the direction is not capturing a random relationship between entity vectors but a distinct attribute-related relationship. We suggest the following comparisons for significance testing: % One can do this by conducting several significance tests to analyze entity cosine similarities with the resulting attribute association bias direction for statistical significance. 

\begin{itemize}
    \item \textit{Cosine similarities between the bias direction and entities in opposing entity sets \(A\) and \(B\)}: This test determines if the two sets have significantly different relationships with the bias direction. Suppose the cosine similarities for \(A\) and \(B\) are not statistically significant. In that case, one can assume that the direction is not capturing a significant attribute difference between the two entity groups or that there is no attribute difference to be captured.

    \item \textit{Cosine similarity between the bias direction and entities in \(A\) and \(B\) versus the entities' cosine similarity with a randomly-sampled vector}: This test examines if the entity sets have a statistically significant relationship with the calculated bias direction versus a random direction. If testing the attribute cosine similarity and random cosine similarity results in significance, one can assume that the attribute direction vector captures behavior that does not occur randomly.

    \item \textit{Cosine similarity of entity vectors from \(A\) (or \(B\)) with the bias direction and that of random vectors with the bias direction}: This test builds upon the previous test to determine if the relationship between the entity sets and the bias direction is significant. It specifies that the entity has a significant relationship with the bias direction in comparison to random entities and the direction. This further validates that the relationship between the entities and computed bias direction is not random. 
\end{itemize}
% We can test this relationship by comparing cosine similarities with the bias direction between the entity sets \(A\) and \(B\). This test determines if the opposing entity sets have significantly different relationships with the bias direction. Suppose the cosine similarity between the two groups is not statistically significant. In that case, one can assume that the direction is not capturing a significant attribute difference between the two entity groups or that there is no attribute difference to be captured. One should also test the cosine similarity of the entity sets with the bias direction against their cosine similarity with a direction created via random sampling. If testing the attribute cosine similarity and random cosine similarity results in significance, one can assume that the attribute direction vector captures behavior that does not occur randomly. The third test analyzes the significance between the cosine similarity of an entity group with the bias direction and that of random vectors with the bias direction. This further tests that the relationship between the entity and the bias direction is not random. If all three tests show statistical significance, one can assume that the bias direction captures a non-random relationship between the two attribute-defining entity sets.

All three tests must show statistical significance for one to determine that the bias direction captures a non-random relationship between the two attribute-defining entity sets and the calculated bias directions. Given the number of statistical tests conducted, one may wish to leverage a p-value with the Bonferroni correction or other techniques to account for multiple tests for significance \cite{shani2011evaluating}.

\subsubsection{Bias Evaluation Metrics}
Permutation tests can be used to determine the significance of the attribute association bias evaluation metrics \cite{shani2011evaluating, caliskan2017}. When evaluating EAA metrics, one can test for the significance of the entity-specific metric, GEAA, and the entity-difference metric, DEAA. A significant GEAA means a biased relationship exists between the entity group and the defined attribute. A significant DEAA represents a significant difference in the level of attribute association bias between the two entity test sets.
For R-RIPA, permutation tests can be used to determine if specific entity set association attribute bias is significant. One can test for a significant difference in attribute association bias between entity test sets by comparing the two populations' cosine similarity scores with the bias direction leveraged when calculating R-RIPA. An alternative method is to calculate R-RIPA for smaller samples of the test entity sets and apply the Wilcoxon rank-sum test or similar \cite{shani2011evaluating}.

\subsubsection{Classification Results}
Analyzing the results of classification scenarios should account for how the scenario was scoped and the goal of the bias evaluation. For example, suppose a practitioner is analyzing the potential for attribute association bias in a downstream model that uses learned item embeddings from a recommendation system. In that case, they should first measure overall accuracy to determine if the embeddings relay the attribute correctly. If the practitioner is also concerned with unfair levels of attribute association bias across items in specific stereotype groups, one could leverage classification fairness metrics to compare performance across specified groups, such as, demographic parity or equalized odds\cite{mehrabi2021survey}. We refer to \citet{mehrabi2021survey} for an overview of classification fairness or bias metrics in such scenarios.

\section{Case Study}
\label{sec:casestudy}
In this section, we present an industry case study of how we leveraged our framework to analyze attribute association bias in a podcast LFR system. We discuss how we implemented our framework to scope the attribute for bias measurement, identify and implement appropriate measurement methods, and flag potentially embedded bias in our LFR outputs. 
% We will discuss the scope of the evaluation to provide context for decisions related to implementation. We also provide high-level implementation details as a reference guide for practitioners wishing to leverage this framework in their system. Finally, we detail how we framed and chose metrics for our classification scenarios to flag the potential for downstream introduction or reinforcement of attribute association bias. We do not provide specific implementation details for flagging significant bias directions or bias metric results due to the details we provide in the methodology section.

\subsection{Scope}
\subsubsection{Deciding On An Attribute}
Our case study evaluates user gender attribute association bias in item embeddings stemming from LFR models. This disaggregated evaluation targets understanding how recommendations of specific pieces or types of content may vary according to a user's gender. Beyond the scope of this work, user gender bias has been largely unexplored concerning recommendation algorithms due to the difficulty in obtaining the user's gender for analysis. As mentioned in \S\ref{sec:related}, gender bias has been primarily explored relating to creator or provider gender in media recommendation -- e.g., distributional fairness of artist gender in music recommendations \cite{melchiorre2021investigating,epps2020artist} and distribution of author gender in book recommendations \cite{ekstrand_gender_dist}. Outside of academic settings, gender bias has also been studied in several industry recommender systems. For example, the audit by \citet{geyik_fairness-aware_2019} examining ranking on LinkedIn is an example of evaluating the \textit{rank fairness} according to gender in a  recommendation system in production. \citet{chen2018investigating} evaluated how gender affects \textit{rank performance} on job sites, such as Indeed and Monster. \citet{dash2021umpire} conducted a large scale audit on \textit{ranking bias} in relation to gender for Amazon recommendations. These industry case studies showcase how gender bias previously studied in more qualitative settings can be quantified and observed in recommendation settings.

The influence of gender on media preferences is a well-researched phenomenon outside of recommendation systems. \citet{wuhr2017tears} studied gendered genre preference in movies, finding that women preferred romance and drama while men preferred genres including science fiction, action, and adventure. Gendered genre preferences have also been shown for digital games; \citet{lange2021time} conducted a study on 484 female and male participants comparing actual and stereotyped gender preferences in digital game genres. Gendered genre preferences have also been found in studies concerning the motivation for reading books and music listening \cite{thelwall2019reader, millar2008selective, berkers2012gendered}. Our case study contributes to both gender bias industry research and academic observations of gendered media preferences, specifically gender bias in podcast recommendations. 

Our case study provides a quantitative deep dive into biased listening behavior previously observed in qualitative academic studies. These studies provided guidance for grouping our entity vectors to implement our analysis methods; we  decided that exploring user gender attribute association bias would provide a good case study for experimenting with our framework and contributing novel insight into this area of research.
For example, \citet{boling2018undisclosed} found listeners of true crime podcasts were predominately female and showed three specific motivations. \citet{craig2021podcasting} found that motivations for podcast use in young adults did not significantly change across gender but across genres, signaling a potential change in gender and genre combined. \citet{soto2022just} presented results showing that various demographic parameters, including gender, drove podcast interests in Latina/o/x young people. However, these stereotypes have yet to be researched quantitatively via bias evaluation in the context of recommendation systems.
In this paper, we continue investigating the relationship between gender and genre by analyzing and quantifying potential user gender bias captured in the trained latent space representation from a recommendation model.

\subsubsection{Defining Entity Sets}
After deciding to target user gender attribute association bias, we needed to determine how to group our entities for analysis. We can group male and female users into our attribute-defining entity sets to target user gender. Nevertheless, we must also determine how to frame our entity sets to test for user gender bias given the variety of possible stereotypes associated with user gender. For example, one could frame the analysis to understand if there is user gender attribute association bias regarding creator gender. An Edison report focusing on podcast listening behavior of women found that women would listen to more podcasts if there were more female hosts within the podcast space \cite{lazovick2022women}. Another industry study by AT\&T found that men were likelier to listen to podcasts hosted by men \cite{attright}.

Instead of framing our analysis that way, we chose to target how user gender may become associated with specific \emph{genres} of podcasts. We looked to past research on podcast genre listening behaviors by gender to determine which genres we should define as test entity groups. In particular, \citet{boling2018undisclosed} noted that true crime podcast listeners are more likely to be female than male, and true crime is one of the most popular genres in female listening \cite{lazovick2022women}. In contrast, sports podcasts have been found to have a primarily male listenership \cite{statista_2022}. When observing proprietary data concerning gender share in listenership, we confirmed that these two genres were significantly skewed towards women for true crime and towards men for sports. Given these findings, we explored gender attribute association bias for podcasts labeled as true crime or sports. 

Our podcast vectors were labeled by predetermined podcast genres. These genre labels were defined via self-selection from podcast hosts and behind-the-scenes cataloging of podcasts. Because a podcast can be classified under multiple genres, we required podcasts labeled as true crime not to be labeled as sports and vice versa.

%Due to past research on binary stereotypes in gender in information retrieval literature, we chose to approach stereotypes in a binary sense between feminine and masculine genres \cite{ekstrand_gender_dist, raj2022fire} In respect to genre categories, we leverage predetermined genre specifications for podcasts. These genres are attributed to podcasts via self-selection from podcast hosts as well as behind the scenes cataloguing of podcasts. In all, there are 21 genres available for analysis and podcasts can be classified under multiple genres. Given past research, we focus on differences in bias between true crime and other genres. Additionally, we create groups of genres based on historical gendered listening: stereotypically male, female, or neutral. A genre is assigned to a specific gender group if more than 65\% of historical listeners are of that gender. At the time of this research, the female genre group consisted of Health \& Fitness, True Crime, and Kids \& Family while the male genre group consisted of Sports, Technology, and Leisure.

\subsubsection{Determining Where To Evaluate}
 Recommender systems in industry settings are commonly designed as hybrid recommendation systems consisting of multiple components to create final predictions \cite{burke2007hybrid}. Often these consist of candidate generation and ranking components \cite{amatriain2015recommender}. In many settings, the candidate generation component leverages a LFR algorithm (such as DNN or collaborative filtering) to create user and item embeddings; these embeddings are used by downstream components to further refine the candidate pool via k-nearest neighbors or providing final rankings with rank-specific algorithms \cite{amatriain2015recommender}. Within this type of recommendation system, we hypothesize that attribute association bias can be introduced via LFR components. We test this by measuring attribute association bias from a LFR algorithm and then observing how attribute association bias can be captured via item vectors with classification models. We plan for future iterations of this framework to provide ways to quantify attribute association bias in ranked lists.

We evaluate user gender attribute association bias regarding podcast genre by implementing our framework on an industry hybrid recommendation system. We decided to evaluate the first component of the system as the earliest point in which this type of bias could be introduced into final recommendations. This component is an industry production-level candidate generation model for podcast recommendation; it uses an LFR algorithm to create pools of podcast vectors by user to be ranked for final recommendation lists.
More specifically, candidates are generated via a deep neural network (DNN) recommendation model, a setup commonly used in industry systems  \cite{nazari2020, covington_deep_2016}. It mimics matrix factorization collaborative filtering via a DNN and is trained with candidate sampling and importance weighting to account for potential popularity bias. Model inputs include user features, podcast ids, and binary labels representing positive or negative implicit feedback. Final user and podcast (or item) representation embedding vectors are collected as outputs from the model.

We trained the model with and without using user gender as a feature to understand the counterfactual effects and potential for implicit bias when trained without explicit use of the sensitive feature. This also enabled us to explore use of our framework for creating baselines for mitigation methods, such as removing sensitive attributes from a model. All analysis and training were conducted offline due to the sensitivity of mitigating user gender bias in an online industry system.

\subsubsection{Scoping Evaluation Data}

We created our evaluation data set by randomly sampling 9,500 female and male users to create a final set of 19,000 users. Our podcast vector data set comprised 31,181 English podcasts from the DNN recommendation model. We restricted our analysis of recommendations to users registered in the United States and podcasts created by English speakers. We chose this subset of data to minimize the possibility of location and language confounds, which could potentially affect gender bias measurements due to differences in cultural norms. In the future, it would be interesting to research how gender stereotypes are found as algorithmic bias differently in recommendations concerning the location and language of the users and served content.

\subsection{Implementation}
This section describes implementation of our proposed techniques for this case study. We defined our attribute-defining entity sets, \(F\) and \(M\), as users labeled as female or male. Our test entity sets, \(S\) and \(TC\), were defined as podcasts categorized as sports and true crime respectively. Podcasts in entity set \(S\) are mutually exclusive with those in \(TC\). Based on our scoped evaluation goals, we chose these groupings to observe the effects of user gender attribute association bias on genres of podcast embeddings. In order to test the ability of our methods to flag changes in bias, we implemented a simple mitigation of removing user gender as a feature during model training, resulting in an evaluation of embeddings trained with and without user gender.

\subsubsection{Latent Space Visualizations} 
We apply PCA for dimensionality reduction to visualize embeddings of users and podcasts trained with and without gender. 
We used the 200 male and 200 female users with the greatest cosine similarity with our attribute bias vector. 
We then projected the podcast embeddings onto the vector space learned by PCA applied to the 400 most biased users to observe clustering along the same axis. Since the first principal component was highly correlated with our ``bias'' vector, we can consider it a proxy for gendered meaning in our podcasts. Therefore, for our feature vectors to have a degree of variation explained by this same vector demonstrates the latent representation of gender in feature vectors. 

\begin{figure}
    \includegraphics[scale=0.4]{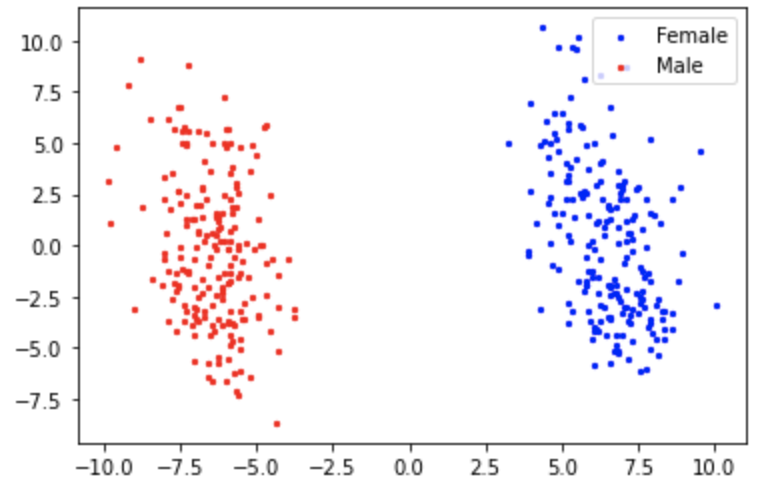}
    \includegraphics[scale=0.4]{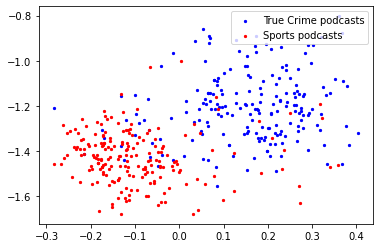}
    \caption{Projection of user embeddings along the first and second PCA components of the 400 most biased users trained with gender (top). Podcasts trained in the same embedding space also show clusters along the same principal components (bottom).}
    \label{fig:pca_proj}
\end{figure}

We replicated this process with our user and podcast embeddings trained without a gender feature. If removing the gender feature removed gendered meaning from the user embeddings, we would expect the same set of 400 ``biased'' users not to be segregated along the first principal component. We used the same set of 400 users and retrained PCA on the ``without gender'' embeddings. 

\begin{figure}
    \includegraphics[scale=0.4]{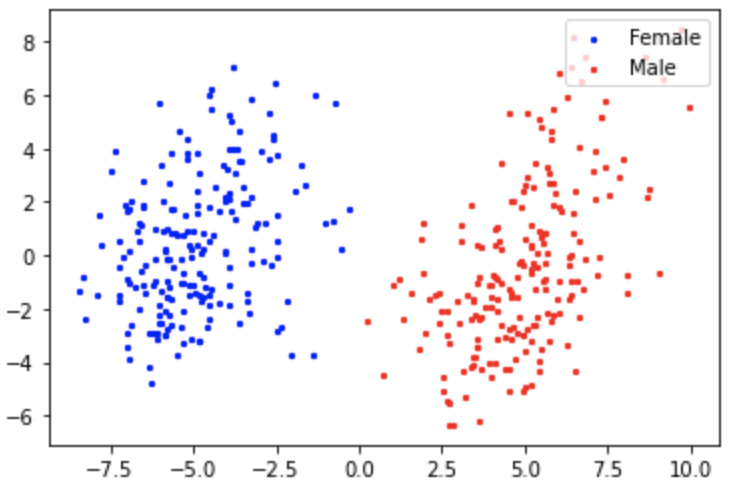}
    \includegraphics[scale=0.4]{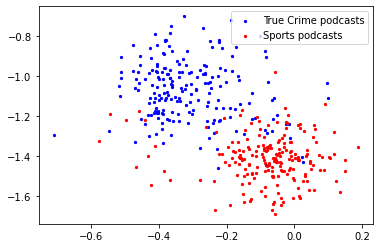}
    \caption{Projection of user embeddings along the first and second PCA components of the 400 most biased users trained without gender (top). Podcasts trained in the same embedding space also show clusters along the same principal components (bottom).}
    \label{fig:pca_proj_ng}
\end{figure}

\subsubsection{Bias Directions}
We used gender-labeled user vectors for each defined approach (centroid, SVC, and PCA) to create the bias directions. These directions were then used to evaluate possible attribute association bias of the test podcast entity sets.

\paragraph{Centroid Difference}  For centroid difference, we simply subtracted the centroid vector of set \(F\) from that of set \(M\).

\paragraph{SVC}  We experimented with three ways to create our SVC model direction based on the entities included during training. The first direction did not consider possible attribute bias of the entities by training the SVC model on a random sample of 6,000 male and 6,000 female users. 
The final attribute layer of this model was the final user gender bias direction. We evaluated the training accuracy to determine if the model successfully captured significant gender meaning. 

We chose to implement multiple levels of perceived bias to allow us to analyze how results are affected by the types of entities leveraged during training. For the following two directions, we defined our entity sets based on their expected level of attribute bias, inspired by \citet{gonen2019lipstick} where they only leveraged word vectors heavily associated with societal gendered stereotypes to create directions. We used the centroid difference direction for identifying users with the most ``biased'' representations, meaning they had the highest cosine similarity with the centroid bias vector. We then trained the SVC model on the 200 and 2500 most ``biased'' male and female user vectors (400 and 5000 user vectors total); we refer to these as CSVC-1 and CSV-2 respectively.

\paragraph{PCA}  When implementing the PCA direction method, we needed a method to create user embedding pairings to capture the user gender attribute since our training data did not explicitly specify user pairings based on user gender. We did so by randomly sampling pairs between male and female user vectors. We did not account for the centroid bias of the users during sampling. The final gender direction vector was the first eigenvector of the resulting principal components of these random user pairings. 

\subsubsection{Bias Evaluation Metrics}
In order to implement our entity attribute association metrics, we defined our two attribute-defining groups and test entity groups as described above. Implementing these metrics represented the association between female users and stereotypically ``female'' podcasts and vice versa with male users and podcasts. We used the gender bias direction vectors described above as our relation vectors for calculating R-RIPA. 

\subsubsection{Classification Scenarios}
We experimented with three classification training scenarios to explore how user gender can affect recommendations differently. We chose to explore how models trained on user vectors in entity sets \(F\) and \(M\) to predict user gender can be repurposed to predict other classification scenarios leveraging only podcast vectors. By doing so, we can evaluate if similar patterns of user gender bias are embedded into item vectors as in user vectors.

In our first classification scenario, we change the classification target variable to predict if most of a podcast's listeners are female or male. We explore the model's performance by analyzing how predicted labels change according to the percentage split between genders. The podcast vectors are used as input into the original models trained to predict user gender from user vectors, with comparison between the gender label output and actual listening splits of the podcast. This scenario is modeled to observe if podcast vectors can relay or reinforce gendered listening behavior.

Second, we use a binary classification model to predict if a user is female or male based on their past podcast listening history. The training data consisted of the centroid of top-3 podcast vectors from each user's listening history. The target variable was the user's gender. We formulated our classification model to observe the potential for podcast vectors to relay user gender bias in downstream recommendation system components. Additionally, in many industry systems, trained recommendation representation embeddings are leveraged in other models to represent content as a feature. If podcast vectors can successfully predict a user's gender, it could signal that these vectors may inadvertently amplify or introduce gender bias into a downstream machine learning model.

Finally, we assess bias at the genre level to determine if genre previously flagged by qualitative studies as experiencing high rates of gendered preference, like true crime and sports podcasts, are more or less affected by the gender bias direction \cite{statista_2022, craig2021podcasting}. We train a binary classification model to predict the ``gender'' of the podcasts in the test entity sets \(S\) and \(TC\). 

\subsection{Flag}
We apply the significance tests proposed in the methodology section to validate the implemented bias directions and metrics. This testing step included three tests of significance for the bias directions and permutation testing to determine the significance of changes in our EAA metric results. 

For the classification scenarios, we tracked precision, recall, and F1-score across true crime and sports podcasts. We set true labels for podcasts based on their associated stereotype, so ``true'' labels for true crime and sports were ``female'' and ``male'' respectively. We used a chi-squared test to determine if results were significantly different when trained with or without user gender as an explicit feature. We compared results between the genre groups with a Mann-Whitney test between the distribution of ``correctly'' predicted podcasts.

When evaluating the gendered listening majority percentage, we observed these metrics at different deciles of gender listenership; this allowed us to provide insight into how attribute association bias may change for item embeddings with higher or lower rates of gendered listening. One can test for significance between deciles with a chi-square test comparing results between two deciles. However, for our analysis, we leveraged this analysis as a way to break down results and were less concerned with the significance between deciles. Understanding differences between these groups would be necessary for more targeted mitigation methods.

\section{Results}
This section presents results from implementing our framework and flagging the existence or change in bias when conducting a simple mitigation method. The goal of this evaluation was to test whether our methodologies were successful in capturing and understanding user gender attribute association bias from our case study's model.

\label{sec:results}
\subsection{Latent Space Visualizations}
Figures~\ref{fig:pca_proj} \& \ref{fig:pca_proj_ng} show latent space visualizations of user and podcast embeddings with and without user gender as a model feature respectively. 
In both cases, our first principal component had a high cosine similarity to the centroid difference (0.89 and 0.91, with gender and without gender as a feature, respectively). We observed both user clusters and podcast clusters emerge along the first principal component for the embeddings trained with and without user gender as a feature. 
When we projected the entity embeddings trained without gender, we observed a similar pattern of clustering as in the case of podcasts trained with gender. 

Both the user and podcast embeddings trained without gender have flipped directionality for the gender direction due to being trained separately from the with-gender embeddings. However, we observe the same relationship of gendered clustering for users along the first principal component in both contexts. Our projections demonstrate that user and podcast embeddings trained without a gender vector still have latent gendered meanings encoded. In the case of podcasts, we observe a weaker separation along this axis, although it is still possible that feature clusters could be derived. 

\subsection{Bias Directions}
When testing the resulting bias directions for significance, we found that all possible direction methods, except for the PCA bias direction, resulted in significant results for our three recommended statistical tests. With the Bonferroni correction, statistical tests were considered significant if p < 0.0033. Beyond the PCA direction, bias directions were significant regardless of if user gender was or was not an explicit attribute used during model training. However, numeric test statistics were greater for our statistical tests when user gender was present during training.

We found that the PCA direction created from random pairings between male and female user vectors failed the test comparing the cosine similarities between the entity and bias direction versus that of random vectors and the bias direction. Since the other two tests were significant, this pointed to the bias direction capturing a significant difference between the attribute entity groups and a significant direction in the space but not a significant relationship between the entity attribute and the bias direction. This result means that the bias direction should not be used for further analysis of gender attribute association bias in the space, since it could easily capture other attribute behavior within the latent space and thus result in inaccurate observations of attribute association bias. It is essential to statistically test one's bias directions since acting on inaccurate results could inadvertently introduce more harm instead of lowering harm in subsequent mitigation.

In addition to our proposed flagging methodology, 
we evaluated the SVC direction based on its model's test accuracy. We split our data into training and test data on an 80-20 split; the results are in Table~\ref{table:svc_perf}. Our first iteration SVC model trained on a random subset of users achieved 99.3\% test accuracy in our with-gender user vectors and 82.2\% test accuracy in our non-gender user vectors. Our mixed-method Centroid-SVC direction (CSVC-1) trained on the 200 ``most biased'' users achieved 96.0\% with-gender test accuracy and 73.6\% non-gender test accuracy. The other mixed-method CSVC direction (CSVC-2) trained on the 2500 ``most biased'' users according to the centroid difference gender direction (CD) achieved 96.4\% with-gender test accuracy and 81.5\% non-gender test accuracy. In order to maintain consistency between test accuracies, we evaluated using the same test set across all trained SVC models. 

\begin{table}
\centering
\begin{tabular}{clcc}
\multicolumn{2}{l}{}                                             & \multicolumn{2}{l}{SVC Accuracy}                                                       \\ 
\cline{3-4}
\multicolumn{2}{c}{SVC Training Data}                            & Train                                     & Test                                       \\ 
\hline
\multirow{2}{*}{SVC}    & {\cellcolor[rgb]{0.898,0.898,0.898}}WG & {\cellcolor[rgb]{0.898,0.898,0.898}}0.992 & {\cellcolor[rgb]{0.898,0.898,0.898}}0.993  \\
                        & NG                                     & 0.829                                     & 0.829                                      \\
\multirow{2}{*}{CSVC-1} & {\cellcolor[rgb]{0.898,0.898,0.898}}WG & {\cellcolor[rgb]{0.898,0.898,0.898}}1.0   & {\cellcolor[rgb]{0.898,0.898,0.898}}0.960  \\
                        & NG                                     & 0.959                                     & 0.736                                      \\
\multirow{2}{*}{CSVC-2} & {\cellcolor[rgb]{0.898,0.898,0.898}}WG & {\cellcolor[rgb]{0.898,0.898,0.898}}1.0   & {\cellcolor[rgb]{0.898,0.898,0.898}}0.964  \\
                        & NG                                     & 0.943                                     & 0.815                                     
\end{tabular}
\caption{Training and test accuracy of three SVC models used to create bias directions: trained on all users (SVC), the 400 most biased users (CSVC-1), and the 5000 most biased users (CSVC-2). User ``bias'' was calculated from the cosine similarity with the centroid direction. Accuracy is shown for when user gender was leveraged during training (WG) and then removed as a simple mitigation technique (NG).}
\label{table:svc_perf}
\end{table}

The significant decrease in test accuracy for CSVC-2 could be attributed to multiple factors, such as a reduction in training data, overfitting to training data, or a reduction in gender explainability for all users via the CD. The difference between with-gender and non-gender test accuracy for CSVC-1 is particularly interesting since we intentionally trained the with-gender and non-gender models on the same users. The reduced ability of these 200 users to accurately predict gender when embeddings are not explicitly trained with user gender as a feature signals that the explicit use of this feature does strengthen the significance of gender in user embeddings within the trained latent space. The perfect training accuracy achieved by training our SVC models on the ``most'' biased users, as found by calculating the cosine similarity with the centroid vector, one can see that the gender centroids do accurately capture gender associations. Additionally, the difference in training and test accuracy for these two models signals that the more stereotypically ``gender biased'' the user is, the more easily they can be linearly separated by an SVC model. The decrease in training and test accuracy when removing gender as a model attribute shows a reduction, but not complete erasure, in gender attribute association in the latent space.

Finally, the reduced accuracy of the CSVC-1 model with no gender may signal that the resulting bias direction may not be best for calculating bias metrics or analyzing attribute association bias. It would be more prudent to leverage bias directions with high levels of significance via testing and high levels of accuracy (when using SVC to create the bias direction).

We calculated the cosine similarity between each possible direction vector to compare the resulting bias directions. The cosine similarities between the different gender directions reflected in with-gender and non-gender embeddings can be found in Table~\ref{table:cosine}. Our comparisons signaled high levels of similarity between all calculated gender directions except for the PCA bias direction. As previously noted, this direction was created by randomly pairing female and male users, finding the difference in their vector directions, and then finding the first eigenvector of their vector differences. Given that this direction is the only one with a low level of cosine similarity, carefully choosing pairs when leveraging this method is essential. We found that randomly pairing users to create a difference vector based on their attribute resulted in our bias direction not accurately capturing potential attribute association bias. The PCA bias direction method should solely be used if the practitioner is confident in their entity pairing methodology to reflect the targeted attribute.

Table~\ref{table:cosine} also shows that the cosine similarity between bias directions from the different methods varied significantly. Even though each direction was significant, this difference demonstrates that each relationship captured is slightly different according to the method used. Additionally, we noticed that these fluctuations decreased when user gender was removed as a model feature. This decrease was expected since user gender was no longer used as a model feature. Additionally, it showcased that the bias directions were capable of relaying implicit, or potentially systematic, bias in the latent space. Given the fluctuations found, we believe it would be responsible for practitioners to explore and test multiple bias directions during analysis to enable more nuanced viewpoints of attribute association bias.

\begin{table*}
\centering
\begin{tabular}{r|l|clllcccc}
\multicolumn{2}{c|}{\multirow{2}{*}{\begin{tabular}[c]{@{}c@{}}Bias \\Direction\end{tabular}}} & \multicolumn{2}{c}{CSVC-1}                                                                                              & \multicolumn{2}{l}{CSVC-2}                                                                                               & \multicolumn{2}{c}{CD}                                                                                              & \multicolumn{2}{c}{PCA}                                                           \\ 
\cline{3-10}
\multicolumn{2}{c|}{}                                                                          & \multicolumn{1}{l}{WG}                                   & NG                                                           & WG                                                        & NG                                                           & \multicolumn{1}{l}{WG}                                   & \multicolumn{1}{l}{NG}                                   & \multicolumn{1}{l}{WG}                & \multicolumn{1}{l}{NG}                    \\ 
\hline
\multirow{2}{*}{SVC}    & WG                                                                   & 0.71                                                     & \multicolumn{1}{c}{-}                                        & \multicolumn{1}{c}{0.75}                                  & \multicolumn{1}{c}{-}                                        & 0.79                                                     & -                                                        & 0.03                                  & -                                         \\
                        & {\cellcolor[rgb]{0.898,0.898,0.898}}NG                               & {\cellcolor[rgb]{0.898,0.898,0.898}}                     & \multicolumn{1}{c}{{\cellcolor[rgb]{0.898,0.898,0.898}}0.40} & \multicolumn{1}{c}{{\cellcolor[rgb]{0.898,0.898,0.898}}-} & \multicolumn{1}{c}{{\cellcolor[rgb]{0.898,0.898,0.898}}0.88} & {\cellcolor[rgb]{0.898,0.898,0.898}}-                    & {\cellcolor[rgb]{0.898,0.898,0.898}}0.89                 & {\cellcolor[rgb]{0.898,0.898,0.898}}- & {\cellcolor[rgb]{0.898,0.898,0.898}}0.03  \\
\multirow{2}{*}{CSVC-1} & WG                                                                   &                                                          &                                                              & \multicolumn{1}{c}{0.99}                                  & \multicolumn{1}{c}{-}                                        & 0.98                                                     & -                                                        & 0.08                                  & -                                         \\
                        & {\cellcolor[rgb]{0.898,0.898,0.898}}NG                               & \multicolumn{1}{l}{{\cellcolor[rgb]{0.898,0.898,0.898}}} & {\cellcolor[rgb]{0.898,0.898,0.898}}                         & \multicolumn{1}{c}{{\cellcolor[rgb]{0.898,0.898,0.898}}}  & \multicolumn{1}{c}{{\cellcolor[rgb]{0.898,0.898,0.898}}0.53} & {\cellcolor[rgb]{0.898,0.898,0.898}}-                    & {\cellcolor[rgb]{0.898,0.898,0.898}}0.38                 & {\cellcolor[rgb]{0.898,0.898,0.898}}- & {\cellcolor[rgb]{0.898,0.898,0.898}}0.11  \\
\multirow{2}{*}{CSVC-2} & \multicolumn{1}{c|}{WG}                                              &                                                          &                                                              &                                                           &                                                              & 0.99                                                     & -                                                        & 0.06                                  & -                                         \\
                        & \multicolumn{1}{c|}{{\cellcolor[rgb]{0.898,0.898,0.898}}NG}          & \multicolumn{1}{l}{{\cellcolor[rgb]{0.898,0.898,0.898}}} & {\cellcolor[rgb]{0.898,0.898,0.898}}                         & {\cellcolor[rgb]{0.898,0.898,0.898}}                      & {\cellcolor[rgb]{0.898,0.898,0.898}}                         & {\cellcolor[rgb]{0.898,0.898,0.898}}                     & {\cellcolor[rgb]{0.898,0.898,0.898}}0.89                 & {\cellcolor[rgb]{0.898,0.898,0.898}}- & {\cellcolor[rgb]{0.898,0.898,0.898}}0.01  \\
\multirow{2}{*}{CD}     & WG                                                                   &                                                          &                                                              &                                                           &                                                              & \multicolumn{1}{l}{}                                     & \multicolumn{1}{l}{}                                     & 0.03                                  & -                                         \\
                        & {\cellcolor[rgb]{0.898,0.898,0.898}}NG                               & \multicolumn{1}{l}{{\cellcolor[rgb]{0.898,0.898,0.898}}} & {\cellcolor[rgb]{0.898,0.898,0.898}}                         & {\cellcolor[rgb]{0.898,0.898,0.898}}                      & {\cellcolor[rgb]{0.898,0.898,0.898}}                         & \multicolumn{1}{l}{{\cellcolor[rgb]{0.898,0.898,0.898}}} & \multicolumn{1}{l}{{\cellcolor[rgb]{0.898,0.898,0.898}}} & {\cellcolor[rgb]{0.898,0.898,0.898}}  & {\cellcolor[rgb]{0.898,0.898,0.898}}0.06 
\end{tabular}
\caption{Cosine similarity for gender direction vectors created through the following methods: SVC model trained on random sample of 12,000 users (SVC), SVC model trained on 400 most ``biased'' users (CSVC-1),  SVC model trained on 5000 most ``biased'' users (CSVC-2), centroid difference (CD), and PCA first eigenvector of the difference between 1000 randomly generated male-female vector pairs (PCA). Similarities were calculated for gender direction vectors created from embeddings trained with gender (WG) and without gender (NG).}
\label{table:cosine}
\end{table*}

\subsection{Bias Amplification Metrics}
\paragraph{EAA, GEAA, and DEAA}  Our results using this set of metrics signaled that our test entity sets of true crime and sports podcast vectors showed significant association attribute bias with their respective user gender.
We found that podcast embeddings trained with and without gender resulted in a significant DEAA score of 612.27 and 480.59, respectively. The calibration effect for with and without gender DEAA was 1.81 and 1.78. The normalization of the calibration effect showcases that the attribute association bias remains highly significant when accounting for the EAA distributions.

EAA metrics successfully flagged a significant change in attribute association bias levels when removing gender. However, significant levels of bias remained. When accounting for the separate GEAA for sports and true crime podcasts, we found that the final DEAA score could be contributed primarily to sports podcasts versus true crime podcasts. When trained with gender, sports podcasts GEAA was -521.34, which was reduced to -406.59 when trained without gender. This decrease of 22\% was greater than the 18.6\% decrease for the true crime podcast test metric.
True crime GEAA was originally 90.93 when trained with gender and reduced to 73.98 after the mitigation. This discrepancy reflects that sports podcasts have significantly higher attribute association bias and are heavily associated with the male attribute-defining entity set of vectors.
This difference also highlights that this simple mitigation method does not equally address gender attribute association bias across groups. Observing the different levels of EAA metrics allow a practitioner to pinpoint which group is more or less affected by the mitigation.

\paragraph{R-RIPA}  We find that R-RIPA also successfully relays attribute association bias; results are in Table~\ref{table:ripa}. 
%When comparing R-RIPA results to EAA results, we found the differential in scores to be higher. This difference could result from EAA reflecting an average of relationships across entities, while R-RIPA captures only one observation for the cosine similarity between the item embedding and bias direction vector. 
When comparing the R-RIPA across bias directions, a couple of results stand out. First, after removing gender, the SVC R-RIPA for true crime podcasts increased, signifying an increase in attribute association bias for true crime podcasts with female users. However, this result is not present for R-RIPA created with the bias directions CSVC-1, CSVC-2, and CD. This difference signals that user gender attribute association bias may have a more nuanced relationship with individual female users that is not fully captured by centroid-based directions. Additionally, we find that true crime podcasts do not experience as significant of a decrease as sports podcasts for R-RIPA calculated with the bias directions CSVC-1, CSVC-2, and CD. This result could signify that removing user gender reduced attribute association bias more heavily for sports podcasts with high levels of attribute association bias as captured by a centroid-related direction.

\paragraph{Metric Comparison}  Unlike the EAA metrics, R-RIPA is at risk of more fluctuation in results depending on the bias direction selected for calculation. As a result, we recommend that practitioners compute R-RIPA only with bias directions that more accurately represent the attribute behavior in the latent space. For example, when leveraging bias directions other than that of SVC (trained on a randomly sampled user set), there is a significant increase in attribute association bias signaled by R-RIPA. This peculiarity could be seen as those bias directions over-reporting bias or SVC under-reporting bias. Additionally, R-RIPA computed with the SVC bias direction is at risk of becoming less accurate as the trained SVC becomes less accurate. It is essential to account for this possibility by implementing permutation testing to determine the significance of R-RIPA results if there is less confidence in the bias direction. In such cases, it may be more prudent to apply the EAA bias metrics instead. 

\begin{table*}[ht]
\setlength{\tabcolsep}{8pt}
\renewcommand{\arraystretch}{1.2}
\centering
\begin{tabular}{lcccccccccc}
                                                             & \multicolumn{10}{c}{Bias Direction Used for Calculating R-RIPA}                                                                       \\
                                                             & \multicolumn{2}{c}{SVC} & \multicolumn{2}{c}{CSVC-1} & \multicolumn{2}{c}{CSVC-2} & \multicolumn{2}{c}{CD} & \multicolumn{2}{c}{PCA}  \\ 
\hhline{~==========}
                                                             & \multicolumn{10}{c}{Model Training Data}                                                                                              \\
                                                             & WG     & NG             & WG     & NG                & WG     & NG                & WG     & NG            & WG     & NG              \\ 
\hline
\multicolumn{11}{c}{Podcast Genre R-RIPA}                                                                                                                                                            \\
\multicolumn{1}{r}{True Crime}                               & 0.132  & 0.182          & 0.118  & 0.115             & 0.137  & 0.131             & 0.138  & 0.129         & 0.012  & 0.054           \\
\rowcolor[rgb]{0.898,0.898,0.898} \multicolumn{1}{r}{Sports} & -0.142 & -0.141         & -0.239 & -0.062            & -0.234 & -0.187            & -0.234 & -0.209        & -0.061 & -0.151          \\
\multicolumn{11}{c}{R-RIPA Summary}                                                                                                                                                                  \\ 
\hline
\multicolumn{1}{r}{Differential}                             & 0.274  & 0.323          & 0.357  & 0.177             & 0.371  & 0.318             & 0.372  & 0.338         & 0.073  & 0.205           \\
\rowcolor[rgb]{0.898,0.898,0.898} \multicolumn{1}{r}{Effect} & 1.686  & 1.768          & 1.787  & 1.339             & 1.812  & 1.763             & 1.808  & 1.784         & 0.494  & 1.430          
\end{tabular}
\caption{R-RIPA results for podcast embeddings (trained with gender (WG) and without gender (NG)) for true crime or sports podcasts leveraging gender directions created from methods described in \S\ref{subsec:biasvectors}. Negative and positive results signify male and female association, respectively.}
\label{table:ripa}
\end{table*}

\subsection{Classification Scenarios}
The classification scenarios we designed allowed us to observe if podcast embeddings used as downstream features resulted in either accurate predictions of user gender engagement or stereotyped predictions of podcasts labeled for our entity test sets.
For each scenario, we evaluated results for podcasts trained with and without user gender as a feature to understand implicit user gender bias in the latent space and how explicit use of the feature amplifies said bias. We used the same SVC classification models trained on user vectors to create gender directions for our analysis: SVC, CSVC-1, and CSVC-2.

\subsubsection{Gendered Podcast Listening}
We analyzed whether these predictions aligned with actual podcast listenership gender percentages. We did this by observing how our SVC models labeled podcasts as ``male" and ``female". We compared these predictions against the podcasts' male and female listenership percentage. In ~\ref{table:flist}, we see the pattern that as podcasts have increasing percentages of male or female listenership, the podcasts are more likely to be classified as ``male" or ``female" podcasts. For example, with the SVC model trained on user embeddings with user gender, we see that when podcasts are in the 50\% decile, they are classified as ``female" 70.8\% of the time, but when female listenership grows to over 70\%, podcasts are labeled as ``female" over 95\% of the time. This classification scenario allows us to see that as engagement becomes more gendered, the podcast entity embeddings become more associated with a specific gender as well.

Interestingly, predictions correlating with female podcast listening became more accurate when the model was not trained with gender. However, this result did not hold for male podcast listening. When the model was trained without gender, the predictions became significantly less accurate when labeling a podcast with higher male engagement as male. Given this change in result, we hypothesized that the semantic embedding of user gender might not precisely represent the female and male binary relationship for podcast vectors but that of male and not male. Understanding how this relationship is embedded into the space would require more in-depth testing with non-binary data, which is out of the scope of this paper but could be an interesting development to explore in future research.

We found that this classification scenario showcased how podcast entity vectors can capture user gender attribute association bias based on the increase in accuracy in predictions as the percentage of listener gender rose. Additionally, the results showed that podcast embeddings associated with male listening experienced a sharper increase in accuracy as the male listener percentage increased. This finding is helpful during evaluation because it flags a difference in behavior within the latent space for podcast embeddings more related to stereotypical male listening.

\begin{table*}[ht]
\centering
\setlength{\tabcolsep}{8pt}
\renewcommand{\arraystretch}{1.5}
\begin{tabular}{cllcccccccccccc}
\multicolumn{3}{l}{}                                                                                                  & \multicolumn{12}{c}{Bias Direction Model}                                                                                                                                                                                                                                                                                                                                                                                                                                                                                                      \\
\multicolumn{3}{l}{}                                                                                                  & \multicolumn{4}{c}{SVC}                                                                                                                                                       & \multicolumn{4}{c}{CSVC-1}                                                                                                                                                    & \multicolumn{4}{c}{CSVC-2}                                                                                                                                                     \\ 
\hhline{~~~============}
\multicolumn{3}{l}{}                                                                                                  & \multicolumn{12}{c}{Model Training Data}                                                                                                                                                                                                                                                                                                                                                                                                                                                                                                       \\
\multicolumn{3}{l}{}                                                                                                  & \multicolumn{2}{c}{WG}                                                                & \multicolumn{2}{c}{NG}                                                                & \multicolumn{2}{c}{WG}                                                                & \multicolumn{2}{c}{NG}                                                                & \multicolumn{2}{c}{WG}                                                                & \multicolumn{2}{c}{NG}                                                                 \\ 
\hhline{~~~============}
\multicolumn{3}{l}{}                                                                                                  & \multicolumn{12}{c}{Predicted Label for Podcast}                                                                                                                                                                                                                                                                                                                                                                                                                                                                                               \\
\multicolumn{3}{l}{}                                                                                                  & \multicolumn{1}{l}{M}                     & \multicolumn{1}{l}{F}                     & M                                         & F                                         & M                                         & F                                         & M                                         & F                                         & M                                         & F                                         & M                                         & F                                          \\ 
\hhline{~~~============}
\multirow{11}{*}{\rotatebox[origin=c]{90}{\% Decile of Listeners by Gender}} & \multirow{5}{*}{\rotatebox[origin=c]{90}{Female}} & 50                                     & 0.292                                     & 0.708                                     & 0.133                                     & 0.867                                     & 0.249                                     & 0.751                                     & 0.137                                     & 0.863                                     & 0.232                                     & 0.768                                     & 0.201                                     & 0.799                                      \\
                                                   &                         & {\cellcolor[rgb]{0.882,0.882,0.882}}60 & {\cellcolor[rgb]{0.882,0.882,0.882}}0.135 & {\cellcolor[rgb]{0.882,0.882,0.882}}0.865 & {\cellcolor[rgb]{0.882,0.882,0.882}}0.051 & {\cellcolor[rgb]{0.882,0.882,0.882}}0.949 & {\cellcolor[rgb]{0.882,0.882,0.882}}0.093 & {\cellcolor[rgb]{0.882,0.882,0.882}}0.907 & {\cellcolor[rgb]{0.882,0.882,0.882}}0.059 & {\cellcolor[rgb]{0.882,0.882,0.882}}0.941 & {\cellcolor[rgb]{0.882,0.882,0.882}}0.093 & {\cellcolor[rgb]{0.882,0.882,0.882}}0.907 & {\cellcolor[rgb]{0.882,0.882,0.882}}0.090 & {\cellcolor[rgb]{0.882,0.882,0.882}}0.910  \\
                                                   &                         & 70                                     & 0.048                                     & 0.952                                     & 0.017                                     & 0.983                                     & 0.027                                     & 0.973                                     & 0.019                                     & 0.981                                     & 0.025                                     & 0.975                                     & 0.034                                     & 0.966                                      \\
                                                   &                         & {\cellcolor[rgb]{0.882,0.882,0.882}}80 & {\cellcolor[rgb]{0.882,0.882,0.882}}0.015 & {\cellcolor[rgb]{0.882,0.882,0.882}}0.985 & {\cellcolor[rgb]{0.882,0.882,0.882}}0.003 & {\cellcolor[rgb]{0.882,0.882,0.882}}0.997 & {\cellcolor[rgb]{0.882,0.882,0.882}}0.008 & {\cellcolor[rgb]{0.882,0.882,0.882}}0.992 & {\cellcolor[rgb]{0.882,0.882,0.882}}0.007 & {\cellcolor[rgb]{0.882,0.882,0.882}}0.993 & {\cellcolor[rgb]{0.882,0.882,0.882}}0.008 & {\cellcolor[rgb]{0.882,0.882,0.882}}0.991 & {\cellcolor[rgb]{0.882,0.882,0.882}}0.010 & {\cellcolor[rgb]{0.882,0.882,0.882}}0.990  \\
                                                   &                         & 90                                     & 0.045                                     & 0.955                                     & 0.034                                     & 0.966                                     & 0.030                                     & 0.970                                     & 0.030                                     & 0.970                                     & 0.042                                     & 0.958                                     & 0.036                                     & 0.958                                      \\
                                                   &                         &                                        & \multicolumn{1}{l}{}                      & \multicolumn{1}{l}{}                      & \multicolumn{1}{l}{}                      & \multicolumn{1}{l}{}                      & \multicolumn{1}{l}{}                      & \multicolumn{1}{l}{}                      & \multicolumn{1}{l}{}                      & \multicolumn{1}{l}{}                      & \multicolumn{1}{l}{}                      & \multicolumn{1}{l}{}                      & \multicolumn{1}{l}{}                      & \multicolumn{1}{l}{}                       \\
                                                   & \multirow{5}{*}{\rotatebox[origin=c]{90}{Male}}   & 50                                     & 0.524                                     & 0.476                                     & 0.303                                     & 0.697                                     & 0.536                                     & 0.464                                     & 0.298                                     & 0.702                                     & 0.529                                     & 0.471                                     & 0.421                                     & 0.579                                      \\
                                                   &                         & {\cellcolor[rgb]{0.882,0.882,0.882}}60 & {\cellcolor[rgb]{0.882,0.882,0.882}}0.715 & {\cellcolor[rgb]{0.882,0.882,0.882}}0.285 & {\cellcolor[rgb]{0.882,0.882,0.882}}0.516 & {\cellcolor[rgb]{0.882,0.882,0.882}}0.484 & {\cellcolor[rgb]{0.882,0.882,0.882}}0.795 & {\cellcolor[rgb]{0.882,0.882,0.882}}0.205 & {\cellcolor[rgb]{0.882,0.882,0.882}}0.515 & {\cellcolor[rgb]{0.882,0.882,0.882}}0.485 & {\cellcolor[rgb]{0.882,0.882,0.882}}0.789 & {\cellcolor[rgb]{0.882,0.882,0.882}}0.211 & {\cellcolor[rgb]{0.882,0.882,0.882}}0.666 & {\cellcolor[rgb]{0.882,0.882,0.882}}0.334  \\
                                                   &                         & 70                                     & 0.868                                     & 0.132                                     & 0.747                                     & 0.253                                     & 0.940                                     & 0.060                                     & 0.727                                     & 0.273                                     & 0.942                                     & 0.058                                     & 0.851                                     & 0.149                                      \\
                                                   &                         & {\cellcolor[rgb]{0.882,0.882,0.882}}80 & {\cellcolor[rgb]{0.882,0.882,0.882}}0.949 & {\cellcolor[rgb]{0.882,0.882,0.882}}0.051 & {\cellcolor[rgb]{0.882,0.882,0.882}}0.903 & {\cellcolor[rgb]{0.882,0.882,0.882}}0.097 & {\cellcolor[rgb]{0.882,0.882,0.882}}0.981 & {\cellcolor[rgb]{0.882,0.882,0.882}}0.019 & {\cellcolor[rgb]{0.882,0.882,0.882}}0.888 & {\cellcolor[rgb]{0.882,0.882,0.882}}0.112 & {\cellcolor[rgb]{0.882,0.882,0.882}}0.981 & {\cellcolor[rgb]{0.882,0.882,0.882}}0.019 & {\cellcolor[rgb]{0.882,0.882,0.882}}0.945 & {\cellcolor[rgb]{0.882,0.882,0.882}}0.055  \\
                                                   &                         & 90                                     & 0.957                                     & 0.043                                     & 0.931                                     & 0.069                                     & 0.973                                     & 0.028                                     & 0.928                                     & 0.072                                     & 0.974                                     & 0.026                                     & 0.956                                     & 0.044                                     
\end{tabular}
\caption{We leveraged our bias direction SVC models to predict gender labels of podcast entity vectors. This table shows how these predicted labels change based on the percent of listeners who are male or female in comparison to the type of model used and if gender was (WG) or was not (NG) leveraged during training.}
\label{table:flist}
\end{table*}

\subsubsection{User Gender from Podcast Listening History}
We designed this scenario to capture the ability of item embeddings to relay sensitive information about users in downstream models. If item embeddings can be used to predict the user-sensitive attribute, as well as the user embedding itself, it can be assumed that the sensitive attribute is entangled within the item embedding.

We found the overall change in test accuracy to be small when leveraging podcast vectors trained with and without access to gender as a feature. Classification test accuracy for with-gender podcast vectors was 0.832, while non-gender podcast vectors achieved a test accuracy of 0.829. When breaking down results by gender, we found the change in test accuracy to be more pronounced. When gender was included as a feature, 17.9\% of female users were classified as male. This percentage reduced to 11.3\% when vectors were trained without access to gender. Alternatively, misclassification for male users increased to 23.7\% when the model was trained without gender versus 15.7\% when trained with user gender as a feature. To better understand the vectors resulting in misclassification, we evaluated the cosine similarity of these vectors against the female, male, female podcast-listening, and male podcast-listening centroids. We found that misclassified podcast vectors showed higher cosine similarity with the opposite gender and gendered listening centroids. 

\subsubsection{Gender Bias by Genre}
Finally, we examine if gender stereotyped genres are more or less likely to be associated with misclassifications of gender if gender is used as a feature or not. This association is evaluated by observing the predicted labels of the sports or true crime podcasts. Results are in Table~\ref{table:pod_perf}.

We found that results for true crime and sports podcasts from SVC and CSVC-2 remained relatively stable when gender was and was not used as a feature during training. When testing for significance, we found that both models did not experience a significant change, with p-values of 0.007 and 0.264, respectively.

However, we found this untrue when testing CSVC-1, which was trained on the 200 ``most gender-biased'' users. Predictions from CSVC-1 showed a significant change in the precision of predicting true crime podcasts as female, with the metric reducing from 0.80 to 0.49. This drop means more sports podcasts were classified as ``female'' instead of ``male.'' One can speculate that this reflects attribute association bias for sports podcasts concerning the 200 ``most gender-biased'' users to have been significantly reduced when removing user gender from the training process. This behavior is also reflected in the significant drop in recall for sports podcasts regarding the CSVC-1 model. In contrast, true crime podcasts experience a slight uptick in the recall, with an increase of 0.81 to 0.84 for CSVC-1 and SVC and CSVC-2 results. 

These results show an imbalanced effect of the chosen mitigation method to remove user gender from training. This assumption is further supported when testing for significance between the genre groups of model performance. When gender was used during training, model performance for predicting the stereotyped gender for a podcast was significantly different for all three classification models. If gender was removed during training, we found that the difference in performance was no longer significant for the SVC model. This difference was not due to the lessening of bias when predicting the gender of a podcast but instead from the classification model predicting more true crime podcasts as ``female.''

\begin{table}
\centering
\begin{tabular}{l|rccc}
\multicolumn{3}{l}{\multirow{3}{*}{}}                                                                               & \multicolumn{2}{c}{Podcast Genre}                                                    \\ 
\cline{4-5}
\multicolumn{3}{l}{}                                                                                                & Sport                                    & True Crime                                \\ 
\cline{4-5}
\multicolumn{3}{l}{}                                                                                                & \multicolumn{2}{c}{Precision}                                                        \\
\multirow{20}{*}{\rotatebox[origin=c]{90}{Bias Direction Model}} & \multirow{2}{*}{SVC}    & WG                                     & 0.96                                     & 0.74                                      \\
                                                 &                         & {\cellcolor[rgb]{0.902,0.902,0.902}}NG & {\cellcolor[rgb]{0.902,0.902,0.902}}0.97 & {\cellcolor[rgb]{0.902,0.902,0.902}}0.74  \\
                                                 & \multirow{2}{*}{CSVC-1} & WG                                     & 0.94                                     & 0.80                                      \\
                                                 &                         & {\cellcolor[rgb]{0.902,0.902,0.902}}NG & {\cellcolor[rgb]{0.902,0.902,0.902}}0.94 & {\cellcolor[rgb]{0.902,0.902,0.902}}0.49  \\
                                                 & \multirow{2}{*}{CSVC-2} & WG                                     & 0.95                                     & 0.80                                      \\
                                                 &                         & {\cellcolor[rgb]{0.902,0.902,0.902}}NG & {\cellcolor[rgb]{0.902,0.902,0.902}}0.95 & {\cellcolor[rgb]{0.902,0.902,0.902}}0.79  \\ 
\cline{4-5}
                                                 & \multicolumn{2}{c}{}                                             & \multicolumn{2}{c}{Recall}                                                           \\
                                                 & \multirow{2}{*}{SVC}    & WG                                     & 0.91                                     & 0.87                                      \\
                                                 &                         & {\cellcolor[rgb]{0.878,0.878,0.878}}NG & {\cellcolor[rgb]{0.878,0.878,0.878}}0.91 & {\cellcolor[rgb]{0.878,0.878,0.878}}0.89  \\
                                                 & \multirow{2}{*}{CSVC-1} & WG                                     & 0.94                                     & 0.81                                      \\
                                                 &                         & {\cellcolor[rgb]{0.878,0.878,0.878}}NG & {\cellcolor[rgb]{0.878,0.878,0.878}}0.74 & {\cellcolor[rgb]{0.878,0.878,0.878}}0.84  \\
                                                 & \multirow{2}{*}{CSVC-2} & WG                                     & 0.94                                     & 0.83                                      \\
                                                 &                         & {\cellcolor[rgb]{0.878,0.878,0.878}}NG & {\cellcolor[rgb]{0.878,0.878,0.878}}0.93 & {\cellcolor[rgb]{0.878,0.878,0.878}}0.84  \\ 
\cline{4-5}
                                                 & \multicolumn{2}{c}{}                                             & \multicolumn{2}{c}{F1-Score}                                                         \\
                                                 & \multirow{2}{*}{SVC}    & WG                                     & 0.93                                     & 0.80                                      \\
                                                 &                         & {\cellcolor[rgb]{0.878,0.878,0.878}}NG & {\cellcolor[rgb]{0.878,0.878,0.878}}0.94 & {\cellcolor[rgb]{0.878,0.878,0.878}}0.81  \\
                                                 & \multirow{2}{*}{CSVC-1} & WG                                     & 0.94                                     & 0.81                                      \\
                                                 &                         & {\cellcolor[rgb]{0.878,0.878,0.878}}NG & {\cellcolor[rgb]{0.878,0.878,0.878}}0.83 & {\cellcolor[rgb]{0.878,0.878,0.878}}0.62  \\
                                                 & \multirow{2}{*}{CSVC-2} & WG                                     & 0.94                                     & 0.82                                      \\
                                                 &                         & {\cellcolor[rgb]{0.878,0.878,0.878}}NG & {\cellcolor[rgb]{0.878,0.878,0.878}}0.94 & {\cellcolor[rgb]{0.878,0.878,0.878}}0.81 
\end{tabular}
\caption{Classification performance scores when classifying sport podcasts as ``male'' or true crime podcasts as ``female'' when leveraging SVC models used to create bias directions. Acronym descriptions can be found in \S\ref{table:svc_perf}.}
\label{table:pod_perf}
\end{table}

\section{Discussion}
\label{sec:discussion}
Our case study applied our proposed framework to observe how attribute association bias changes when removing user gender as a feature in an industry LFR model. 
\citet{gonen2019lipstick} suggested the capability for user gender bias to be systematic bias embedded within the latent space, thus making it difficult for simple mitigation techniques to address the core issue; they noted ``a systematic bias found in the embeddings, which is independent of the gender direction.'' Given this independence, debiasing methods grounded in removing the gender direction were found to be ``superficial'' fixes. Systematic bias in our case study, similar to that found in word embeddings, would result in significant attribute association bias even when user gender is not included as a model feature.

Like the results by \citet{gonen2019lipstick}, our case study suggests that gender stereotypes can become implicitly embedded in the representations of both users and items, as supported by the persistence of this bias after removing user gender as a feature during model training. We found that removing user gender as a feature resulted in a statistically significant decrease in levels of attribute association bias, but significant attribute association bias still remained. The presence of such implicit attribute association bias signals the potential for systematic gender bias when using LFR models and representations, or potentially recommendation algorithms in general, for serving podcasts to users. This finding is not surprising given previous research detailing the highly gendered nature of podcast listening \cite{boling2018undisclosed, craig2021podcasting, soto2022just}. Our findings based on our framework demonstrate that, similar to \citet{gonen2019lipstick}, known systematic bias can be found and quantified in recommendations. Given this, it is essential for practitioners to audit for attribute association bias when systematic bias is a known factor in their recommendation scenario, such as podcast recommendations.

Additionally, our finding means that more straightforward measures, such as removing gender as a feature during training, can also be seen as a ``superficial'' attempt to entirely remove gender bias from the representation space. Our results concerning the relatively small changes in our metrics and classification outputs show that relying on feature removal in recommendations will not fully mitigate user gender bias entirely from the user or item vectors. 
Additionally, the findings demonstrate that implicit attribute association bias can occur in LFR models, signaling that this type of bias may need to be accounted for in representation learning beyond NLP and image processing.

In light of this, we did find that removing user gender during training of a deep LFR does reduce the inequality of gender association between male and female users to their stereotypical genres and content. In the case of podcast listening, it is unlikely that gender bias can be removed entirely due to the systematic bias found in listening trends showcased by users and found in external research. However, removing gender as a distinct feature could be a first step in reducing the overall amplification of gendered associations reflected in the relationships between user and item vectors within the trained latent space.

Our findings supporting the possibility for systematic bias to occur as attribute association bias in LFR outputs leads us to a more challenging question for the research community: when is it appropriate to mitigate systematic bias? If it is found that implicit attribute association bias improves user experience, how should one reduce the risk of representative harm? In some cases, like ours, stereotyped behavior is typical, and some may argue that it is beneficial for providing valuable content recommendations to users. In the case of user gender bias, the harm lies in the model potentially reinforcing stereotypes by driving users towards gendered listening habits they might otherwise not partake in. It is possible to monitor levels of attribute association bias over time to flag increasing bias in the latent space. Nevertheless, when do levels of reinforcing and implicit bias become harmful? Both the research and practitioner community would benefit from more exploration of how to approach setting baselines for managing representative harms in recommendations. 

\section{Limitations \& Future Work}
\label{sec:limits}
Our most noticeable limitation when implementing these techniques was the lack of distinct and well-labeled user pairings for metric calculation, a common occurrence in recommendation settings. We demonstrated methods to overcome this limitation, but this work could be further refined in the future to avoid possibly introducing more bias into the evaluation via practitioner-defined entity pairing techniques. In the future, specific to this case study, we plan to explore counterfactual user vectors according to gender to create distinct pairs. Counterfactual user pairings would isolate the feature within the latent space and potentially reduce attributing spurious relationships between users solely to gender differences. It is important to note that this workaround is only available for models trained with entity attributes. This limitation would remain when evaluating the implicit or systematic bias of an LFR algorithm.

Another limitation that could be addressed with future work is exploring gender bias in a non-binary manner. In general, algorithmic bias research is done on binary groups because it is inherently easier to measure biased relationships between two groups rather than multiple groups. Methods for measuring bias for multiple groups could be further developed to capture multi-group relationships. Instead of creating novel metrics, researchers could also expand the analysis to compare more than two groups. However, this can quickly create an overwhelming amount of comparisons as the number of groups for evaluation grows.

Finally, this evaluation was performed on a proprietary industry system for a type of media known for listening patterns highly related to the listener's gender. We would like to conduct these evaluation techniques on public data sets (such as Last.FM) and other recommendation algorithms to understand if our methodology performs well when levels of attribute association bias are not as distinct as those found within our podcast recommendations case study.

\section{Conclusion}
\label{sec:end}
Our framework provides a clear path in uncovering potentially harmful stereotyped relationships resulting from attribute association bias resulting from an LFR model. In showcasing our techniques on an industry case study, we found that our proposed methodologies successfully measured and flagged attribute association bias. Additionally, we uncovered clear advantages and disadvantages for our proposed methods to help practitioners choose the appropriate techniques for their scoped evaluations. 

The success of our methodologies in uncovering attribute association bias highlights the importance of understanding how stereotypical relationships can become embedded into trained recommendation latent spaces. For example, our ability to predict user gender from podcast vectors demonstrates how leveraging these vectors as attributes in downstream models can introduce implicit user gender bias in subsequent outputs, even if owners of downstream models intentionally remove user gender as a training feature. The ability for listening history to predict user gender showcases that user gender bias is embedded within the podcast vectors, meaning their use can inherently introduce gender bias into other modeling systems. Understanding this type of representation bias becomes increasingly crucial in industry recommendation systems where embeddings are used across models owned by different teams. 
%For example, if a team audits and mitigates its model for user gender bias but leverages said podcast vectors as a feature, any unrelated models leveraging said feature could be unknowingly reintroduced to user gender bias. 
If attribute association bias is left unchecked in hybrid recommendation scenarios, teams are at risk of amplifying systematic representation harms resulting from providing stereotyped recommendations for stakeholders.

 We also demonstrate that classifiers can serve as valuable tools for uncovering behaviors of bias in representation learning outside of NLP, thus opening the doors for future work leveraging innovative evaluation techniques across representation learning disciplines. 
 Based on our results, training classification models to highlight potential attribute association bias helped showcase more nuanced behavior that may be lost when solely using other metrics. The ability to design specific classification scenarios enabled us to observe different ways bias could be captured and how item embeddings assumed to be associated with specific attributes can display varying behavior. We observed nuanced results such as increased prediction accuracy for podcast embeddings associated with high levels of female listening when user gender was not a feature, which would otherwise require more complex metrics and methodologies to highlight how behavior between entities differs when attribute association bias is present.

Similar to findings by \citet{basta2019evaluating}, our results support the notion that capturing and understanding the behavior of gender bias in more implicitly biased recommendation vector embeddings is a complicated and nuanced task, requiring further analysis beyond our results showcased in this paper.
We hope our evaluation framework serves as a building block for future research addressing representative harms and attribute association bias in recommendation systems.

\bibliographystyle{ACM-Reference-Format}
\bibliography{refs}
\end{document}